# Boris Computational Spintronics – High Performance Multi-Mesh Magnetic and Spin Transport Modelling Software

Serban Lepadatu[*]

*Jeremiah Horrocks Institute for Mathematics, Physics and Astronomy, University of Central Lancashire, Preston PR1 2HE, U.K.*

Abstract

This work discusses the design and testing of a new computational spintronics research software. Boris is a comprehensive multi-physics open-source software, combining micromagnetics modelling capabilities with drift-diffusion spin transport modelling and heat flow solver in multi-material structures. A multi-mesh paradigm is employed, allowing modelling of complex multi-layered structures with independent discretization and arbitrary relative positioning between different computational meshes. Implemented micromagnetics models include not only ferromagnetic materials modelling, but also two-sublattice models, allowing simulations of antiferromagnetic and ferrimagnetic materials, fully integrated in the multi-mesh and multi-material design approach. High computational performance is an important design consideration in Boris, and all computational routines can be executed on GPUs, in addition to CPUs. In particular a modified 3D convolution algorithm is used to compute the demagnetizing field on the GPU, termed pipelined convolution, and benchmark comparisons with existing GPU-accelerated software Mumax3 have shown performance improvements up to twice faster.

[*] SLepadatu@uclan.ac.uk

# 1. Overview

Micromagnetics is a field of study concerned with understanding magnetization processes on the continuum scale, and is an invaluable tool in interpreting experimental results, designing spintronics devices, testing analytical methods, and predicting new effects. Existing micromagnetics software include open-source finite difference packages OOMMF [1], Mumax3 [2], and Fidimag [3]. A number of other micromagnetics packages are also available, including finite element/boundary element methods, both open-source and commercial, with a review given in Ref. [4].

**Figure 1** – Overview of computational information flow for Boris. Different computational mesh types may be configured, including ferromagnetic, antiferromagnetic, ferrimagnetic, normal metal, and insulator meshes. Each mesh has several computational modules available, including the Transport and Heat solvers. Supermeshes are the smallest rectangles encompassing all the individual meshes of same type, with specific computational modules available. The information generated is used by an assigned magnetization dynamics equation (LLG/LLB) to evolve the magnetization data in the individual magnetic meshes.

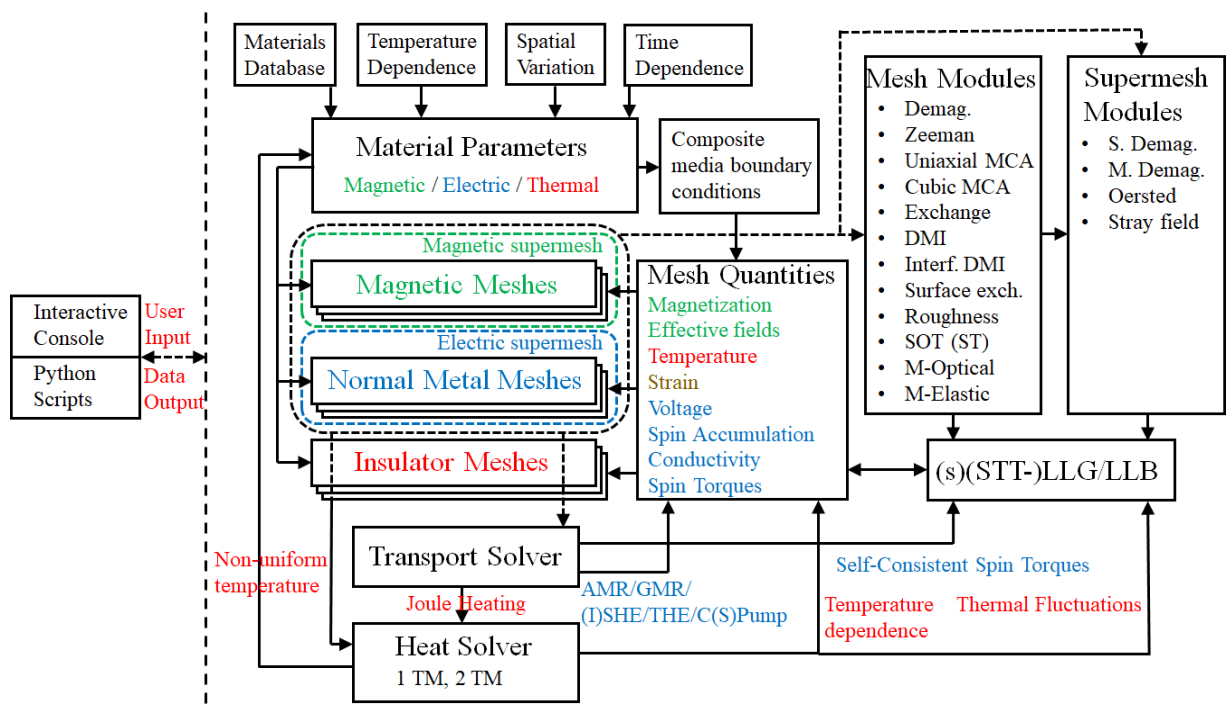

Boris is a micromagnetics-oriented multi-physics research software. In contrast to existing finite difference packages it is specifically designed as a multi-mesh and multi-material software. Arbitrary geometries can be handled, where long-range interactions such as the magnetostatic interaction and Oersted field are calculated across all relevant computational meshes, and short-range interactions between neighbouring meshes are treated using appropriate composite media boundary conditions. An overview is given in Figure 1. Magnetization dynamics are computed using the Landau-Lifshitz-Gilbert (LLG) equation [5], or the Landau-Lifshitz-Bloch (LLB) equation [6], either of which may be augmented by thermal fluctuations (stochastic versions) [7,8], Zhang-Li spin transfer torques (STT) [9], interfacial STT (ISTT) [10], spin-orbit torques (SOT) due to the spin-Hall effect (SHE) [11], Slonczewski spin torques [12], or spin torques computed self-consistently using a spin transport solver. The spin transport solver is based on a drift-diffusion model with circuit theory boundary conditions [13-15], and self-consistently calculates charge currents, spin currents, and spin accumulations in multi-layer structures [16]. In addition to obtaining spin torques self-consistently several effects may be computed, including anisotropic magneto-resistance (AMR), current perpendicular to plane giant magneto-resistance (CPP-GMR) [14], SHE and inverse SHE (ISHE) [17], spin pumping [18], charge pumping and topological Hall effect [19,20]. A heat solver is also available, allowing calculation of heat flow in response to ambient conditions as well as sources and sinks. An important source of heat is due to Joule heating from a current density calculated using the transport solver. This allows inclusion of temperature-dependent effects in the magnetization dynamics, including AMR-generated magnonic spin-Seebeck effect [21]. Another heat source is due to ultrafast laser pulses, and a two-temperature model (2TM) is included to allow simulations of ultrafast demagnetization and recovery processes [22]. Additionally a two-sublattice model is implemented, allowing simulations of antiferromagnetic and ferrimagnetic materials, fully integrated within the multi-mesh computational paradigm, allowing for example simulations with exchange bias. All parameters appearing in the working equations are available as user-controllable material parameters and may be assigned a temperature dependence, spatial variation, and time dependence; several spatial variation generators are available, including Voronoi tessellations, as well as user-defined dependences through mathematical equations or data files.

The computational meshes can be sized and discretized independently. One of the most difficult interactions to compute across several independent computational meshes is the magnetostatic interaction. A newly developed method, termed multi-layered convolution [23], allows computation of demagnetizing fields for multiple meshes with arbitrary thicknesses, arbitrary relative positioning and spacings, without impacting on the computational performance. Other long-range interactions include the Oersted field, which is computed from the current density obtained using the transport solver, as well as stray field computation from a number of fixed magnetic dipoles. Individual magnetic mesh modules include magneto-crystalline anisotropy (MCA), either uniaxial or cubic, direct exchange interaction, Dzyaloshinsky-Moriya interaction (DMI) [24,25], either bulk or interfacial, surface exchange coupling [26], topographical surface and edge roughness [27].

The software has a modular structure and is open-source [28], facilitating community contributions of new computational modules. An extensive user manual [29] is included, together with many examples of both scripted simulations, as well as pre-configured simulation setups. The software is provided with a graphical user interface for interactive display of simulation data, with user control enabled through a graphical console allowing intuitive and interactive control of simulations. The software may also be controlled using Python scripts which communicate with Boris through network sockets, thus allowing either local or remote user control. The software has been programmed mainly in C++17 and CUDA C, as well as Python. All computational routines can be executed on central processing units (CPU), as well as graphical processing units (GPU) using the CUDA framework [30]. Supported operating systems include Windows 7, Windows 10, and Linux-based distributions; in the current version (2.81) Linux compilations of Boris do not include a graphical interface, only providing a basic text console, however the software is otherwise fully functional and may be conveniently controlled using Python scripts. The code-base size currently consists of ~130k source lines of code (comments and trivial lines excluded), and is contained in ~800 source code files, including a purpose-written object-oriented finite difference vector calculus library for both CPU and GPU computations. External dependencies include FFTW3 [31] and CUDA [30]. Material definitions are made available through an online database of material parameters [32]. The online materials database allows users to contribute new entries through a set of simple built-in protocols described in the manual. Material definitions used in this work are given in the online materials database [32]. Moreover all the simulation scripts and files used to obtain the results presented here have been included in the Boris GitHub repository [28].

## 2. Basic Micromagnetics Modelling

In the continuum approximation, magnetization dynamics may be computed using the LLG equation:

$$\frac{\partial \mathbf{m}}{\partial t} = -\gamma \mathbf{m} \times \mathbf{H}_{eff} + \alpha \mathbf{m} \times \frac{\partial \mathbf{m}}{\partial t} \qquad (1)$$

Here **m** is the normalised magnetization direction, $\gamma = \mu_0 g_{rel} |\gamma_e|$, where $\gamma_e = -g\mu_B/\hbar$ is the electron gyromagnetic ratio and $g_{rel}$ is a relative gyromagnetic factor, α is the Gilbert damping factor [33], and **H**$_{eff}$ is an effective field which includes a number of interactions as additive field contributions. In a basic micromagnetics formulation these include an applied field contribution, the magnetostatic or demagnetizing field interaction, and the direct exchange interaction. Depending on the material simulated a magneto-crystalline anisotropy contribution may be included, either uniaxial or cubic, as well as bulk or interfacial DMI [24,25]. Equations for these contributions implemented in Boris are given in Appendix A. A number of evaluation methods are available for the magnetization dynamics equations. These are the fixed step methods Euler (1$^{st}$ order), trapezoidal Euler (2$^{nd}$ order), and Runge-Kuta (RK4 - 4$^{th}$ order). Adaptive time-step methods are the adaptive Heun (2$^{nd}$ order), the multi-step Adams-Bashforth-Moulton (2$^{nd}$ order), Runge-Kutta-Bogacki-Shampine (RK23 – 3$^{rd}$ order with embedded 2$^{nd}$ order error estimator), Runge-Kutta-Fehlberg (RKF45 – 4$^{th}$ order with embedded 5$^{th}$ order error estimator), Runge-Kutta-Cash-Karp (RKCK45 – 4$^{th}$ order with embedded 5$^{th}$ order error estimator), and Runge-Kutta-Dormand-Prince (RKDP54 – 5$^{th}$ order with embedded 4$^{th}$ order error estimator). For static problems a steepest descent solver is available using Barzilai-Borwein stepsize selection formulas [34,35].

A widely used test for the validity and accuracy of LLG solvers is the µMAG Standard Problem #4 [36]. Here the magnetization response to a magnetic field is computed for a Ni$_{80}$Fe$_{20}$ rectangle with dimensions 500 nm × 125 nm × 3 nm, starting from a relaxed S-state. This is used as a test for correct implementation of the LLG equation and associated effective field terms (demagnetizing field, exchange interaction, and Zeeman term) as any errors result in significant deviations of the magnetization time dependence from accepted solutions. The results for the specified Field 1 (µ$_0$H$_x$ = -24.6 mT, µ$_0$H$_y$ = 4.3 mT, µ$_0$H$_z$ = 0.0 mT) are shown in Figure 2, compared with results obtained using OOMMF [1]. Excellent agreement throughout the switching process is obtained, with overall R$^2$ measure between the two data sets of 0.999. A cellsize of 5 nm was used here which is consistent with the exchange length in Ni$_{80}$Fe$_{20}$, $l_{ex} = \sqrt{2A/\mu_0 M_S^2} \cong 5.7$ nm, however Boris has been extensively tested using this problem with cellsize values down to 1 nm, thus including both 2D and 3D modes. Similarly the specified Field 2 ($\mu_0 H_x = -35.5$ mT, $\mu_0 H_y = -6.3$ mT, $\mu_0 H_z = 0.0$ mT) was also tested. The results in Figure 2 were computed using the RK4 method with a fixed time-step of 500 fs, however all the implemented evaluation methods were successfully tested using this problem, both for CPU and GPU computations in single and double floating point precision.

**Figure 2** – Magnetization response computed for µMAG Standard Problem #4 using Field 1 specification, showing the normalized components of magnetization, compared to the magnetization response computed in OOMMF. Overall $R^2$ measure of 0.999 was obtained.

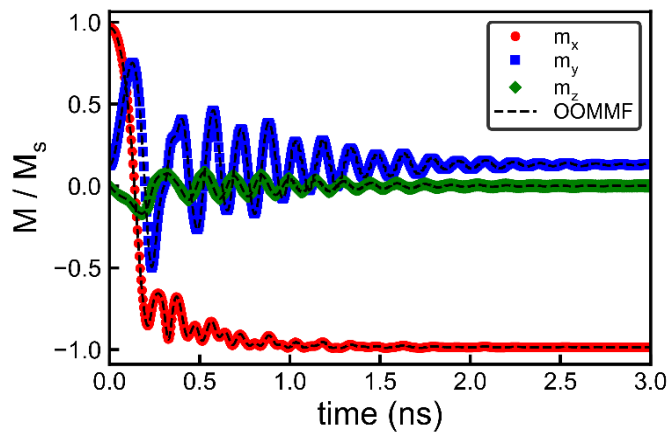

A further test which requires a more advanced external field stimulus consists in computing the spin wave dispersion as described in Ref. [37]. Here a Ni$_{80}$Fe$_{20}$ magnonic waveguide track with 1 µm length, 50 nm width and 1 nm thickness is used, and spin waves are excited using a field pulse given by: H(t) = H$_e$ sinc(k$_C$(x-x$_0$)) sinc(k$_C$(y-x$_0$)) sinc(2πf$_C$(t-t$_0$)). Boris has a provision for input stimuli specification using mathematical formulas, simultaneously allowing spatial and temporal dependence. Here the excitation field amplitude was set to H$_e$ = 400 kA/m, frequency cutoff f$_c$ = 500 GHz, and wave-vector cutoff k$_C$ = 2π×0.1255 rad/nm, as specified in Ref. [37]. Using the Nyquist criterion a time sampling interval of 1 ps, and spatial sampling interval of 4 nm along the 1 µm long track were used. The excitation was applied in the centre of the track for a duration of 2t$_0$, with a temporal sinc pulse centre t$_0$ = 200 ps. Three spin wave geometries are possible, depending on the direction of the bias field, namely i) backward volume for bias field along the length, ii) forward volume for bias field along the thickness, and iii) surface spin waves for bias field along the width. The wave-vector direction for this problem in all cases is along the length of the track, which is determined by the spatial sampling direction. Results for the backward volume are shown in Figure 3 for a damping value α = 0.01, where a bias field $H_0$ = 804 kA/m was used, with the excitation field pulse applied along the width. The spin wave dispersion was obtained using a 2D Fourier transform from the y component of magnetization. The discretization cellsize was set to 1 nm × 2 nm × 1 nm, with periodic boundary conditions [38] used along the length only, and the RK4 method was used with a 50 fs time-step. This results in excellent agreement between the computed spin wave dispersion and the analytical dotted lines given by $w = w_n + 2\gamma A k^2 / \mu_0 M_S$; here $w_n$ are the resonance frequencies obtained at k = 0 rad/m. Similar tests were performed for the two remaining spin wave geometries.

**Figure 3** – Spin wave dispersion spectrum computed for the problem specified in Ref. [37], using a damping of 0.01. The dotted lines are obtained from the formula $w = w_n + 2\gamma A k^2 / \mu_0 M_S$, with $w_n$ being the resonance frequencies obtained at k = 0 rad/m.

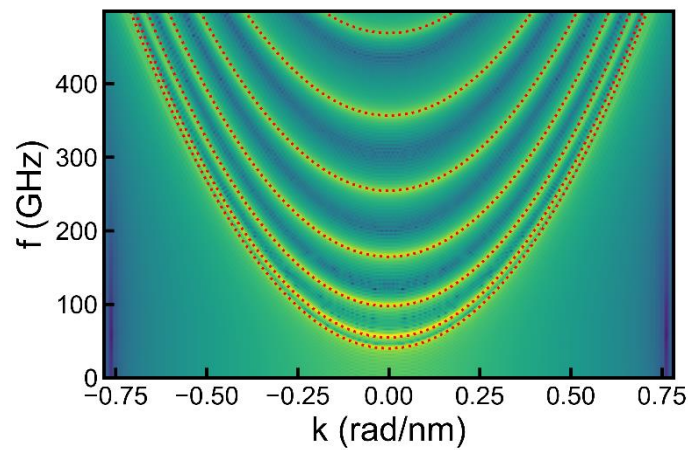

The LLG equation may be modified to include STT, and in particular Boris implements Zhang-Li STT [9,39], given by:

$$\frac{\partial \mathbf{m}}{\partial t} = -\gamma \mathbf{m} \times \mathbf{H} + \alpha \mathbf{m} \times \frac{\partial \mathbf{m}}{\partial t} + (\mathbf{u}.\nabla)\mathbf{m} - \beta \mathbf{m} \times (\mathbf{u}.\nabla)\mathbf{m} \qquad (2)$$

The spin-drift velocity **u** is given by:

$$\mathbf{u} = \mathbf{J}\frac{Pg\mu_B}{2eM_s}\frac{1}{1+\beta^2}, \tag{3}$$

where **J** is the charge current density, *P* is the current spin polarisation, and β is the non-adiabaticity parameter. The LLG-STT equation is widely used for studying the effect of bulk STT on magnetization textures, including transverse domain walls, Bloch and Néel domain walls, vortices and skyrmions. In particular domain wall velocity dependence on spin drift velocity may be computed, including simulation of Walker breakdown phenomenon. Since this is a very common type of computation Boris implements a moving domain wall algorithm, which allows efficient simulation of domain wall movement using a finite track length. End magnetic charges are removed using the stray field computed from magnetic dipoles at each end, and spin waves are absorbed by freezing the magnetization spins at the track ends. The domain wall is kept centred in the track and any domain wall displacement is recorded in a dedicated output parameter. This algorithm was tested previously [40], and it is straightforward to verify the expected relation v/u = β/α [41], with v being the domain wall velocity far from Walker breakdown. Here we show another test of the LLG-STT equation, based on the µMAG Standard Problem #5 [42]. A permalloy rectangle with dimensions 100 nm × 100 nm × 10 nm is initialized with a vortex structure – Figure 4(a) – and a constant current density resulting in a spin drift velocity u = -72.35 m/s is applied for a range of non-adiabaticity parameter values. Results for β = 0.1 are shown in Figure 4(b), using a cubic cellsize of 2.5 nm, plotting the x and y components of magnetization as a function of time. Excellent agreement with results computed in OOMMF is obtained, with overall $R^2$ measure between the two data sets of 0.999. Similar successful tests were performed for the remaining specified values of β = 0, 0.05 and 0.5.

**Figure 4** – Vortex dynamics computed for μMAG Standard Problem #5 for $\beta = 0.1$. (a) Relaxed starting vortex state, also showing the fitted spatial dependence of non-adiabaticity parameter computed using the spin transport drift-diffusion solver. (b) Vortex dynamics are shown for the LLG-STT solver with constant non-adiabaticity, compared to results computed in OOMMF. Overall $R^2$ measure of 0.999 was obtained. Results obtained using the self-consistent bulk spin torque obtained from the drift-diffusion model, where β is no longer constant but varies due to in-plane spin diffusion, are also shown for comparison.

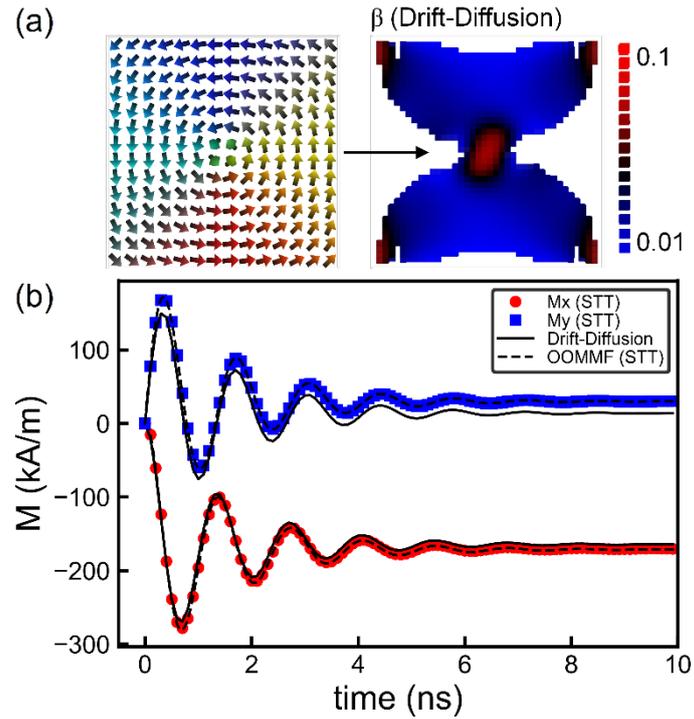

## 3. Multi-Mesh Micromagnetics Modelling

Due to progress in experimental magnetism increasingly devices are composed of complex multi-layered structures, including multi-layered stacks used to study skyrmions [43-46], and synthetic anti-ferromagnetic structures [47-49]. Such multi-layered structures, which cannot be discretized effectively using a single uniform finite difference mesh, are difficult to study using software packages which only implement a single computational mesh without introducing approximations or using unnecessarily small cellsize values. In Boris a multi-mesh paradigm has been adopted from the outset, allowing computations using multiple meshes which can be arbitrarily positioned relative to each other, and with independent discretization cellsize values. Thus whilst still benefitting from computationally efficient finite difference discretization, multi-layered structures commonly found in experimental studies may be simulated without compromising accuracy or computational speed. This is accomplished using a new multi-layered convolution algorithm introduced in Boris [23], used to compute demagnetizing fields for a collection of finite difference computational meshes. For a collection of meshes $V_k$ ($k = 1, …, n$), the convolution sum may be written as:

$$\mathbf{H}(\mathbf{r}'_{kl}) = -\sum_{\substack{i=1,...,n \\ \mathbf{r}_{ij} \in V_i}} \mathbf{N}(\mathbf{r}'_{kl} - \mathbf{r}_{ij}, \mathbf{h}_k, \mathbf{h}_i) \mathbf{M}(\mathbf{r}_{ij}), \quad k = 1,...,n; \quad \mathbf{r}'_{kl} \in V_k \quad (4)$$

Here $\mathbf{r}_{ij}$ is the cell-centred position vector of cell $j$ in mesh $i$ ($i = 1, …, n$), and $\mathbf{N}$ are inter and intra-mesh demagnetizing tensors generalized from the Newell et al. formulas [50] in Ref. [23]. With a single computational mesh the usual approach to efficiently evaluate the convolution sum is to use the convolution theorem, which involves computing the forward Fourier transform of the magnetization, multiplying point-by-point with the Fourier transform of the demagnetizing tensor (kernel) in the transform space, and finally taking the inverse Fourier transform to obtain the demagnetizing field. With multiple input meshes a similar approach may be taken to evaluate Equation (4), with summation of inter-mesh contributions moved to the transform space. Full details, including validation tests, are given in Ref. [23]. Here we extend this algorithm to use periodic boundary conditions based on the multiple images method [51], and demonstrate the use of multi-layered convolution by simulating the hysteresis loop in a [$Co_{90}Fe_{10}$ (4.9 nm) / Ru (0.6 nm) / $Co_{90}Fe_{10}$ (2.9 nm) / Ru (0.6 nm)]$_{10}$ synthetic ferrimagnetic structure. Here the layers preferentially align antiparallel due to RKKY interaction [52-54], as well as due to the magnetostatic field interaction between the layers. For two magnetic layers separated by a metallic spacer, the surface exchange energy density and effective field are given as:

$$\varepsilon = -\frac{J_1}{\Delta}\mathbf{m}_i.\mathbf{m}_j - \frac{J_2}{\Delta}(\mathbf{m}_i.\mathbf{m}_j)^2$$

$$\mathbf{H}_i = \frac{J_1}{\mu_0 M_S \Delta}\mathbf{m}_j + \frac{2J_2}{\mu_0 M_S \Delta}(\mathbf{m}_i.\mathbf{m}_j)\mathbf{m}_j$$

(5)

**Figure 5** – Hysteresis loop in a synthetic ferrimagnetic 10-repetition multi-layered stack of [$Co_{90}Fe_{10}$ (4.9 nm) / Ru (0.6 nm) / $Co_{90}Fe_{10}$ (2.9 nm) / Ru (0.6 nm)]$_{10}$.

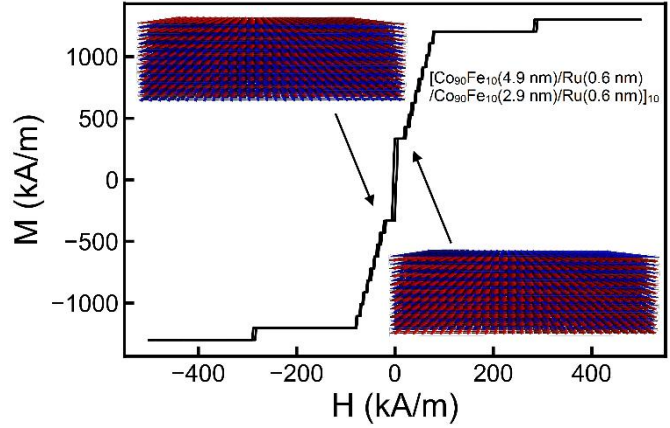

Here $J_1$ and $J_2$ are the bilinear and biquadratic surface exchange coupling constants respectively, with coupled magnetic moment directions given by $\mathbf{m}_i$ and $\mathbf{m}_j$, and $\Delta$ is the thickness of the ferromagnetic layer for which the $\mathbf{H}_i$ effective field contribution is calculated. The simulated structure is shown in the inset to Figure 5, where we've used material parameters determined experimentally in Ref. [47], also available in the materials database [32], and in particular $J_1 = -1$ mJ/m$^2$ with no biquadratic contribution. The simulated stack uses in-plane periodic boundary conditions for a 300 nm$^2$ simulated area, and an in-plane cellsize of 3 nm. A polycrystalline structure has been generated using in-plane Voronoi tessellation with 20 nm crystallites, where the uniaxial anisotropy easy axis varies randomly by ±20° around the x axis between the different crystallites. Results are shown in Figure 5 with the field applied along the x axis direction. With a large external field the magnetization in all the layers aligns along the field. Reduction of the field results in sequential switching of the 10 thinner layers against the field direction, thus reducing the total energy as both the surface exchange and magnetostatic interactions result in preferentially anti-parallel alignment. As the field direction is reversed all the thicker $Co_{90}Fe_{10}$ layers switch at once, with the thinner $Co_{90}Fe_{10}$ layer switching against the field due to the strong antiferromagnetic surface exchange interaction; further increasing the field results in sequential switching of the 10 thinner layers towards the applied field direction. It should be noted that such a structure is very difficult to simulate using a single uniform finite difference computational mesh without introducing approximations, such as rounding the layer thicknesses, which become increasingly inaccurate as the number of repetitions increases. Even with the layer thickness rounded, for example to $Co_{90}Fe_{10}$ (5 nm) / Ru (1 nm) / $Co_{90}Fe_{10}$ (3 nm), a discretization cellsize of 1 nm is still required along the z direction. With the multi-layered convolution algorithm in Boris each layer can be considered as a 2D mesh, rendering such simulations relatively trivial.

Further extensions to the micromagnetics model include topographical roughness and staircase corrections for the demagnetizing field, detailed previously [27] and tested experimentally [55-57], as well as magneto-elastic contributions [58,59]. Finally, all the material parameters included in simulations may be assigned spatial and temporal dependences either through user-supplied mathematical formulas, input data files, or built-in generators. This allows simulations using polycrystalline or granular structures, as well as material defects and impurities.

# 4. Transport Solver

Inclusion of spin torques in modern micromagnetics solvers is an essential requirement, allowing modelling the effect of spin transfer torques on domain walls and skyrmions, spin-torque nano-oscillators [60] and magnetic random-access memories [61]. In the simplest case a uniform current density may be used, with the LLG equation augmented with appropriate spin torque terms. More advanced solvers also allow for non-uniform current densities, thus enabling simulations of structures with non-constant cross-sectional area. In Boris the current density may be computed self-consistently for any given geometry and multi-layered structure without having to import a computed current density, using the successive over-relaxation method [62].

**Figure 6** – Anisotropic magneto-resistance (AMR) in a 320 nm × 160 nm × 10 nm $Ni_{80}Fe_{20}$ ellipse. (a) Simulation geometry showing the computed current density, and (b) AMR loops computed for different angles to the ellipse long axis.

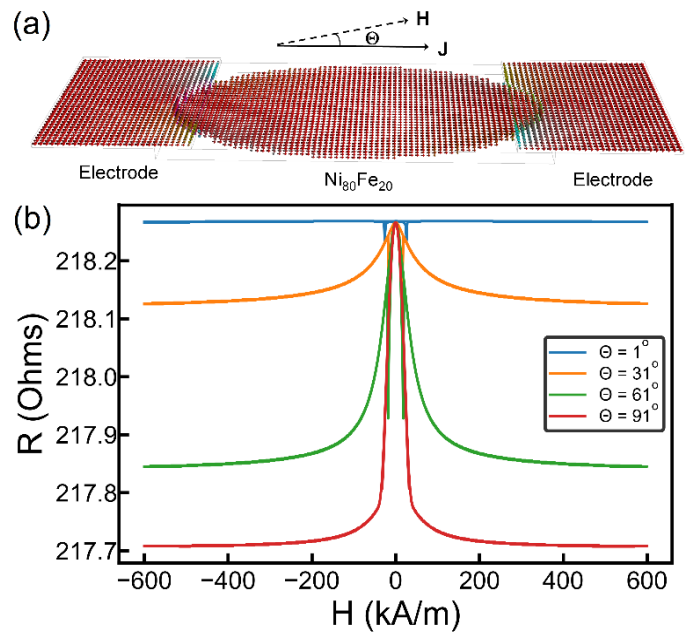

For charge transport only this is given by $\mathbf{J}_c = \sigma\mathbf{E}$, where $\sigma$ is the electrical conductivity and $\mathbf{E} = -\nabla V$ is the electric field obtained from the electrical potential V. Results in Figure 4 made use of the LLG-STT equation with a uniform current density. Here we further show computations with non-uniform current densities, and in particular we compute the AMR in a $Ni_{80}Fe_{20}$ ellipse where a current density is generated by applying a potential drop across two 5 nm thick metallic contacts, which are included as separate computational meshes with Ru material and Dirichlet boundary conditions set at the ends. The contacts are placed on top of the ellipse, with the simulated geometry given in Figure 6(a) showing the computed current density. Composite media boundaries between the ellipse and contacts are treated using continuity of flux (current density) and electrical potential. The AMR effect is included as σ = $\sigma_0$ / (1 + $rd^2$), where $\sigma_0$ is the material base conductivity, r is the AMR ratio taken as 0.02 for $Ni_{80}Fe_{20}$, and d = $\mathbf{J_C}\cdot\mathbf{M}$ / $|\mathbf{J_C}||\mathbf{M}|$. The results are shown in Figure 6(b) where the resistance is obtained as the potential drop across the entire simulated structure divided by the total current flowing into the circuit ground. The AMR loops show the typical behaviour expected for longitudinal and transverse AMR loops, and it is noteworthy the computed resistance change is significantly lower than the input AMR parameter. This is mostly due to the inclusion of constant resistance of the simulated electrical contacts, but also due to non-uniformity of the current density. Test simulations with uniform current density and potential drop applied directly across a single ferromagnetic mesh reproduce the input AMR ratio accurately.

An additional benefit of self-consistently computing current densities, an Oersted field can then be generated from it. This avoids having to compute the Oersted field externally and then importing it into the program, which apart from constant current densities is not trivial. Moreover internal computation of the Oersted field allows simulations with time-dependent Oersted field, for example due to time-dependent current densities. Boris computes the Oersted field from the current density by evaluating the convolution sum with an Oersted tensor, using the formulas given in Ref. [63].

Additionally Boris also allows computation of spin transport based on the drift-diffusion model [13,14], augmented with circuit theory boundary conditions [15]. The full system of equations implemented is shown below.

$$\mathbf{J}_C = \sigma\mathbf{E} + \beta_D D_e \frac{e}{\mu_B}(\nabla\mathbf{S})\mathbf{m} + \theta_{SHA} D_e \frac{e}{\mu_B}\nabla\times\mathbf{S} + P\sigma\frac{\hbar}{2e}\mathbf{E}^\sigma - P\frac{\sigma^2\hbar}{e^2 n}\mathbf{E}\times\mathbf{B}^\sigma$$

$$\mathbf{J}_S = -\frac{\mu_B}{e}P\sigma\mathbf{E}\otimes\mathbf{m} - D_e\nabla\mathbf{S} + \theta_{SHA}\frac{\mu_B}{e}\boldsymbol{\varepsilon}\sigma\mathbf{E} + \qquad (6)$$

$$\frac{\hbar\mu_B\sigma}{2e^2}\sum_i \mathbf{e}_i \otimes(\dot{\mathbf{m}}\times\partial_i\mathbf{m}) + \frac{\hbar\mu_B\sigma^2}{e^3 n}(\mathbf{z}\times\mathbf{E})\otimes(\partial_x\mathbf{m}\times\partial_y\mathbf{m})$$

The charge current density now additionally includes contributions due to i) current perpendicular to plane giant magneto-resistance (CPP-GMR), where $\beta_D$ is the diffusion spin polarization, $D_e$ is the electron diffusion constant and **S** is the spin accumulation. ii) ISHE where $\theta_{SHA}$ is the intrinsic spin Hall angle, iii) charge pumping, and iv) topological Hall effect, where $E_i^\sigma = (\dot{\mathbf{m}} \times \partial_i \mathbf{m}) \cdot \mathbf{m}$ and $\mathbf{B}^\sigma = \mathbf{z}(\partial_x \mathbf{m} \times \partial_y \mathbf{m}) \cdot \mathbf{m}$. Here $\mathbf{E}^\sigma$ and $\mathbf{B}^\sigma$ are the directions of the emergent electric field due to charge pumping, and emergent magnetic field due to topological Hall effect respectively [19,20]. The spin current density tensor, where $\mathbf{J}_{Sij}$ indicates the flow of the *j* component of spin polarisation in the direction *i*, includes contributions due to i) drift, ii) diffusion, iii) SHE where **ε** is the rank 3 unit antisymmetric tensor, iv) charge pumping, and v) topological Hall effect where *n* is the itinerant electron density. The spin accumulation obeys the following equation of motion, where $\lambda_{sf}$ is the spin flip length, $\lambda_J$ is the exchange rotation length, and $\lambda_\varphi$ is the spin dephasing length:

$$\frac{\partial \mathbf{S}}{\partial t} = -\nabla \cdot \mathbf{J}_S - D_e \left( \frac{\mathbf{S}}{\lambda_{sf}^2} + \frac{\mathbf{S} \times \mathbf{m}}{\lambda_J^2} + \frac{\mathbf{m} \times (\mathbf{S} \times \mathbf{m})}{\lambda_\varphi^2} \right) \tag{7}$$

Solving for the spin accumulation allows computation of bulk spin torques, which may be included as an additional torque term in the LLG equation, as:

$$\mathbf{T}_S = -\frac{D_e}{\lambda_J^2} \mathbf{m} \times \mathbf{S} - \frac{D_e}{\lambda_\varphi^2} \mathbf{m} \times (\mathbf{m} \times \mathbf{S}) \tag{8}$$

It may be shown that under the assumption of negligible in-plane spin diffusion this expression is equivalent to Zhang-Li STTs as given in Equation (2) [9,16,64], where the non-adiabaticity parameter is constant and given by $\beta \cong \lambda_J^2 / \lambda_{sf}^2$ in the limit of long spin dephasing length and long domain walls. The assumption of negligible in-plane spin diffusion breaks down for rapidly varying magnetization textures such as vortices and skyrmions and this can lead to spatially varying and enhanced non-adiabaticity. For example it is known that vortex domain walls have a significantly larger non-adiabaticity compared to transverse domain walls [65], arising mainly due to in-plane spin diffusion at large magnetization gradients, with contributions due to charge pumping and topological Hall effect also recognized [66]. Whilst it may still be possible to use the simple LLG-STT formulation of Equation (2), the correct value of non-adiabaticity must be used when vortex domain walls are present, and this may be computed using the drift-diffusion model. This is shown in Figure 4(a), where for $Ni_{80}Fe_{20}$ the relation $\beta \cong \lambda_J^2 / \lambda_{sf}^2$ gives $\beta = 0.04$ expected for a transverse domain wall with $\lambda_{sf} = 10$ nm and $\lambda_J = 2$ nm. For the vortex domain wall in Figure 4(a) however a much higher maximum value of 0.1 results for $\lambda_\varphi = 2.1$ nm, obtained by fitting the spin torque in Equation (8) to the STT in Equation (2) with $\beta$ as a spatially varying fitting parameter. Thus $\beta$ is no longer a constant, but has a spatial dependence with the maximum value reached at the vortex core as seen in Figure 4(a). The µMAG Standard Problem #5 is repeated again, but this time the LLG equation is used with the spin torque from Equation (8), i.e. the spin accumulation is solved at every time step to self-consistently compute the spin torque. The results are shown in Figure 4(b) where a good agreement is obtained with the LLG-STT equation, despite the very different methods used to solve the problem.

At non-magnetic (N) / ferromagnetic (F) composite media boundaries the following conditions are applied, obtained from circuit theory using the spin mixing conductance $G^{\uparrow\downarrow}$ and interface conductances for majority and minority carriers, $G^\uparrow$ and $G^\downarrow$:

$$\begin{aligned}
\mathbf{J}_C.\mathbf{n}\big|_N &= \mathbf{J}_C.\mathbf{n}\big|_F = -(G^\uparrow + G^\downarrow)\Delta V + (G^\uparrow - G^\downarrow)\Delta \mathbf{V}_S.\mathbf{m} \\
\mathbf{J}_S.\mathbf{n}\big|_N - \mathbf{J}_S.\mathbf{n}\big|_F &= \frac{2\mu_B}{e}\left[\mathrm{Re}\{G^{\uparrow\downarrow}\}\mathbf{m}\times(\mathbf{m}\times\Delta\mathbf{V}_S) + \mathrm{Im}\{G^{\uparrow\downarrow}\}\mathbf{m}\times\Delta\mathbf{V}_S\right] \\
\mathbf{J}_S.\mathbf{n}\big|_F &= \frac{\mu_B}{e}\left[-(G^\uparrow + G^\downarrow)(\Delta\mathbf{V}_S.\mathbf{m})\mathbf{m} + (G^\uparrow - G^\downarrow)\Delta V \mathbf{m}\right]
\end{aligned} \qquad (9)$$

Interfacial spin torques are obtained as ($h_F$ is the discretization cellsize of the F layer in the direction normal to the composite media boundary):

$$\mathbf{T}_S = \frac{g\mu_B}{eh_F}\left[\mathrm{Re}\{G^{\uparrow\downarrow}\}\mathbf{m}\times(\mathbf{m}\times\Delta\mathbf{V}_S) + \mathrm{Im}\{G^{\uparrow\downarrow}\}\mathbf{m}\times\Delta\mathbf{V}_S\right] \qquad (10)$$

Spin pumping may also be included on the N side of Equation (9) as:

$$\mathbf{J}_S^{pump} = \frac{\mu_B \hbar}{e^2}\left[\mathrm{Re}\{G^{\uparrow\downarrow}\}\mathbf{m}\times\frac{\partial\mathbf{m}}{\partial t} + \mathrm{Im}\{G^{\uparrow\downarrow}\}\frac{\partial\mathbf{m}}{\partial t}\right] \qquad (11)$$

This results in a damping-like torque in Equation (10), reproducing the expected enhancement in effective magnetization damping [16]. As shown previously, when a heavy metal (HM) / F bilayer is simulated with the SHE enabled in the HM layer, the expected damping-like and field-like SOTs are obtained from Equation (10) [16,67]. Moreover when a spin accumulation is generated at magnetization gradients, such as a skyrmion, the resulting imbalance in spin accumulation either side of the HM/FM interface generates vertical spin currents which leads to an additional type of interfacial spin torque, termed interfacial STT (ISTT) [10]. In many cases it is sufficient to run simulations with direct expressions for spin torques (e.g. STT, SOT) augmenting the LLG equation, however the drift-diffusion spin transport solver is still useful for calculating the strength of these spin torques in the first place from spin transport parameters. Here we further verify the spin transport solver reproduces the expected spin torques in a spin valve structure shown in Figure 7(a). In a macrospin approximation the total spin torque exerted on the free layer is given by a combination of Slonczewski and field-like spin torques as [68,69]:

$$\mathbf{T}_S = \frac{\mu_B}{e} \frac{J}{d} \eta(\theta) [\mathbf{m} \times (\mathbf{m} \times \mathbf{p}) + r\mathbf{m} \times \mathbf{p}]$$

$$\eta(\theta) = \frac{q_+}{A + B\cos(\theta)} + \frac{q_-}{A - B\cos(\theta)}$$

(12)

Here $\mathbf{p}$ is the polarization from the fixed layer, set to $\mathbf{p} = -\hat{\mathbf{x}}$. By varying the angle in the uniformly magnetized free layer the angular dependence $\eta(\theta)$ in Equation (12) is obtained by fitting the spin torque computed self-consistently in Equation (10). The results are shown in Figure 7(b), where we obtain $q_+$ = 4.94, $q_-$ = -0.05, $A$ = 5.85, $B$ = 3.83 and $r$ = 0.19. It should be noted that whilst a good agreement is obtained between Equation (10) and Equation (12) for the angular dependence, the model above is strictly applicable for a macrospin only. During switching of the free layer the magnetization is no longer uniform resulting in non-negligible spin diffusion effects. Whilst the switching times computed with Equation (10) and Equation (12) respectively are approximately the same for the geometry in Figure 7(a), the exact magnetization dynamics of the free layer differ between them. In this case the self-consistent spin torque in Equation (10) is a more accurate description, capturing the non-local nature of the spin torques.

**Figure 7** – Spin torques in a CPP-GMR spin valve structure for a current density ~$10^{12}$ A/m$^2$. (a) Spin-valve geometry with dimensions 160 nm × 80 nm, 10 nm fixed layer thickness, 5 nm free layer thickness, 2 nm spacer thickness and contacts of 20 nm thickness. (b) Computed spin torques in the free layer for uniform magnetization as a function of in-plane angle, with fitted Slonczewski and field-like spin torques.

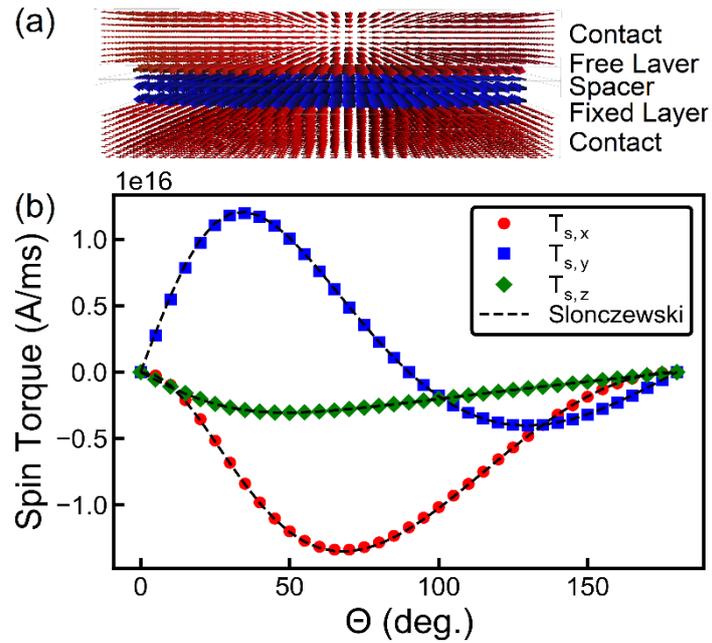

# 5. Heat Solver

Material parameters used in simulations may be assigned temperature dependences, either using a data file, or with a user-supplied mathematical equation. This is particularly useful for computations where the temperature can change during the simulation. In this case the (stochastic) Landau-Lifshitz-Bloch (LLB) equation [8] is used, given by:

$$\frac{\partial \mathbf{M}}{\partial t} = -\frac{\gamma}{1+\tilde{\alpha}_\perp^2}\mathbf{M}\times\mathbf{H} - \frac{\tilde{\alpha}_\perp \gamma}{1+\tilde{\alpha}_\perp^2}\frac{1}{|\mathbf{M}|}\mathbf{M}\times\left(\mathbf{M}\times\left(\mathbf{H}+\mathbf{H}_{thermal}\right)\right) + \frac{\gamma\tilde{\alpha}_\parallel}{|\mathbf{M}|}(\mathbf{M}.\mathbf{H})\mathbf{M} + \mathbf{\eta}_{thermal} \quad (13)$$

Here for T < $T_C$ ($T_C$ is the Curie temperature), $\alpha_\perp = \alpha(1-T/3T_C)$, $\alpha_\parallel = \alpha 2T/3T_C$ and $\tilde{\alpha}_\perp = \alpha_\perp/m$, $\tilde{\alpha}_\parallel = \alpha_\parallel/m$, where $m$ is the magnetization length normalised to its zero temperature value, i.e. $m = |\mathbf{M}|/M_S^0$. For T > $T_C$ $\alpha_\perp = \alpha_\parallel = 2T/3T_C$. The effective field **H** must be complemented by a longitudinal susceptibility field given by:

$$\mathbf{H}_l = \begin{cases} \left(1-\frac{m^2}{m_e^2}\right)\frac{\mathbf{m}}{2\mu_0\tilde{\chi}_\parallel}, & T \leq T_C \\ -\frac{\mathbf{m}}{\mu_0\tilde{\chi}_\parallel}, & T > T_C \end{cases} \quad (14)$$

The field and temperature-dependent equilibrium magnetization, $m_e$, is obtained from the Curie-Weiss law as:

$$m_e(T) = B\left[m_e\frac{3T_C}{T} + \frac{\mu\mu_0 H_{ext}}{k_B T}\right], \quad (15)$$

where B(x) = coth(x) – 1/x is the Langevin function, and μ is the atomic moment. The relative longitudinal susceptibility is given below (units T$^{-1}$), where $\chi_\parallel$ is the longitudinal susceptibility (unitless):

$$\tilde{\chi}_\parallel(T) = \frac{\mu}{k_B T}\frac{B'(x)}{1-B'(x)(3T_C/T)} = \chi_\parallel(T)/\mu_0 M_S^0, \quad with \quad x = m_e 3T_C/T \quad (16)$$

Further, we have the temperature dependences $M_S(T) = M_S^0 m_e(T)$, and $A(T) = A_0 m_e^2(T)$ for the exchange stiffness, although these can be adjusted depending on the simulation requirements. The components of the thermal field, **H**$_{thermal}$, and torque, **η**$_{thermal}$, follow Gaussian distributions with no correlations, zero mean and standard deviations given respectively by:

$$H_\sigma = \frac{1}{\alpha_\perp} \sqrt{\frac{2k_B T(\alpha_\perp - \alpha_\parallel)}{\gamma \mu_0 M_S^0 V \Delta t}}$$

$$\eta_\sigma = \sqrt{\frac{2k_B T \alpha_\parallel \gamma M_S^0}{\mu_0 V \Delta t}}$$

(17)

Here $V$ is the computational cellsize and $\Delta t$ is the integration time-step. In the simplest case the electron temperature in the LLB equation is uniform. More advanced simulations also require non-uniform temperatures, and this calls for implementation of the heat equation. For example Ref. [21] investigated the AMR-induced magnonic spin-Seebeck effect, where Joule heating is included in the heat equation as the heat source $\mathbf{J}^2 / \sigma$ (W/m$^3$). Due to the AMR of a transverse DW the conductivity is higher at the DW, locally resulting in decreased Joule heating. This results in a temperature gradient between the centre of the DW and its boundaries, and moreover when the DW is displaced due to STT the leading edge of the DW experiences a higher temperature compared to the trailing edge. Due to the magnonic spin-Seebeck effect [70] this results in a significant enhancement of the DW velocity up to 15% for realistic material parameters [21]. To reproduce such an effect it is necessary to simultaneously solve both the LLB and heat equations. The heat equation implemented in Boris is shown below, given as the more general two-temperature model [22].

$$C_e \rho \frac{\partial T_e(\mathbf{r},t)}{\partial t} = \nabla . K \nabla T_e(\mathbf{r},t) - G_e(T_e - T_l) + S$$

$$C_l \rho \frac{\partial T_l(\mathbf{r},t)}{\partial t} = G_e(T_e - T_l)$$

(18)

Here $C_e$ and $C_l$ are the electron and lattice specific heat capacities, $\rho$ is the mass density, $K$ is the thermal conductivity, and $G_e$ is the electron-lattice coupling constant, typically of the order $10^{18}$ W/m$^3$K. The simpler one-temperature model may be selected, where effectively $G_e$ is set to zero, and $C_e$ is replaced by a total specific heat capacity, which is sufficient for simulating Joule heating effects.

**Figure 8** – Ultrafast demagnetization and Néel skyrmion creation in a 2 nm thick Co layer on Pt (8 nm) and SiO$_2$ substrate (40 nm). (a) Simulated trilayer structure. (b) State after 800 ps for a high power laser pulse (T$_{max}$ ≅ 2T$_C$, T$_C$ = 500 K) and out-of-plane field of 100 kA/m, showing the z component of magnetization and 5 created skyrmions. (c) |Q| plotted as a function of time for two different pulse strengths (high T$_{max}$ ≅ 2T$_C$, low T$_{max}$ ≅ 1.5T$_C$) and $d$ = 400 nm, $t_R$ = 100 fs. The maximum Co temperature reached for the high power laser pulse is also plotted.

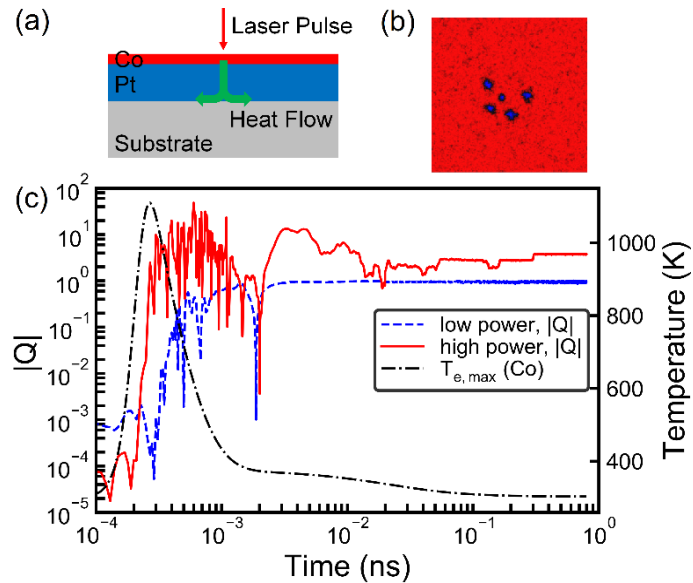

Studies of ultrafast magnetization dynamics have revealed a large difference between electron and spin dynamics on time-scales of the order 1 picosecond and below, explained in terms of a 3-temperature model which includes the electron, spin, and lattice temperatures [71], and later formulated as a microscopic 3-temperature model [72]. This latter approach was shown to be equivalent to an LLB formulation [22] which accounts for the different electron and lattice temperatures on ultra-short time-scales. Within this formulation the photon energy is absorbed by the delocalized electrons, which are coupled to the lattice electrons via the rate equations shown in Equation (18). This allows simulations of ultrafast magnetization due to heating by a laser pulse, by including an appropriate heat source $S$ in Equation (18). Here we shown an example of such a simulation by taking a Gaussian profile for a linearly polarized laser pulse as given below, where $d$ and $t_R$ are full-width at half-maximum values for the spatial and temporal widths.

The geometry simulated is shown in Figure 8(a), where we use a Co (2 nm) / Pt (8 nm) / SiO$_2$ (40 nm) structure, with in-plane dimensions of 512 nm$^2$ and periodic boundary conditions. For the Co layer we also include the interfacial DMI contribution, and uniaxial

$$S = P_0 \exp\left(\frac{-|\mathbf{r}-\mathbf{r}_0|}{d^2/4\ln(2)}\right) \exp\left(\frac{-(t-t_0)^2}{t_R^2/4\ln(2)}\right) \quad (W/m^3) \tag{19}$$

anisotropy with easy axis out of the plane. The two-temperature heat equation is computed for the Co and Pt layers, whilst the one-temperature heat equation is computed for the SiO$_2$ substrate. Continuity of heat flux and temperature is assumed at the interfaces, and Robin boundary conditions are used on the exposed surfaces of the magnetic layer and substrate with ambient temperature set to room temperature. For a high power laser pulse ($P_0 = 4\times10^{21}$ W/m$^3$) with $d = 400$ nm and $t_R = 100$ fs, the computed maximum Co electron temperature is plotted in Figure 8(b), showing ultrafast heating up to $T_{max} \cong 2T_C$ ($T_c = 500$ K), followed by rapid cooling as the electron and lattice temperatures equilibrate. The temperature decays back to room temperature on a longer time scale. For this problem we solve the stochastic LLB in Equation (13) and compute the topological charge (which takes on values ±1 for a single skyrmion) using [73]:

$$Q = \frac{1}{4\pi}\int_A \mathbf{m}\cdot\left(\frac{\partial \mathbf{m}}{\partial x}\times\frac{\partial \mathbf{m}}{\partial y}\right)dxdy \tag{20}$$

Thus by plotting |Q| as a function of time the number of skyrmions present can be monitored. As the magnetization order recovers for T < T$_C$ following ultrafast demagnetization, Néel skyrmions begin to emerge under the action of DMI, as observed experimentally [74]. The mean number of skyrmions formed is dependent on the laser power and follows a Poisson counting distribution as discussed in Ref. [75]. Two examples are shown in Figure 8(c): the low power pulse results in a single skyrmion formed in this case, whilst the high power pulse results in 5 skyrmions formed, with the final state indicated in Figure 8(b). The integration of a multi-layered heat solver with the magnetization dynamics solver is thus a powerful feature, allowing detailed studies with non-uniform and non-constant temperatures.

# 6. Two-Sublattice Model

Recent years have seen an increased interest in antiferromagnetic spintronics [76-78], with the real prospect of antiferromagnetic memories [79] in sight, and applications to terahertz technologies [80]. Thus micromagnetics research software is needed to support future efforts in this area. Following the multi-mesh and multi-material paradigm, Boris has been extended with a two-sublattice model, allowing modelling of antiferromagnetic and ferrimagnetic materials, for example applicable to studies of ferrimagnetic skyrmions [81]. This allows studying not only antiferromagnetic and ferrimagnetic materials devices on their own, but also complex multi-layered devices including both antiferromagnetic and ferromagnetic materials – one obvious application here is to the study of exchange bias [82].

Here we show the 2-sublattice stochastic LLB equation implemented in Boris, based on the LLB equation from Refs. [83,84], applicable for antiferromagnetic, ferrimagnetic, as well as binary ferromagnetic alloys. This is given in Equation (21) in terms of the macroscopic magnetization, where we denote the 2 sublattices as $i = A, B$.

$$\frac{\partial \mathbf{M}_i}{\partial t} = -\tilde{\gamma}_i \mathbf{M}_i \times \mathbf{H}_{eff,i} - \tilde{\gamma}_i \frac{\tilde{\alpha}_{\perp,i}}{M_i} \mathbf{M}_i \times \left( \mathbf{M}_i \times \left( \mathbf{H}_{eff,i} + \mathbf{H}_{th,i} \right) \right)$$
$$+ \gamma_i \frac{\tilde{\alpha}_{\|,i}}{M_i} \left( \mathbf{M}_i \cdot \mathbf{H}_{\|,i} \right) \mathbf{M}_i + \boldsymbol{\eta}_{th,i} \quad (i = A, B) \tag{21}$$

The reduced gyromagnetic ratio is given by $\tilde{\gamma}_i = \gamma_i / (1 + \alpha_{\perp,i}^2)$, and the reduced transverse and longitudinal damping parameters by $\tilde{\alpha}_{\perp(\|),i} = \alpha_{\perp(\|),i} / m_i$, where $m_i(T) = M_i(T)/M_{S,i}^0$, with $M_{S,i}^0$ denoting the zero-temperature saturation magnetization, and $M_i \equiv |\mathbf{M}_i|$. The exchange field now includes not only intra-lattice contributions, but also homogeneous and non-homogeneous inter-lattice contributions given as:

$$\mathbf{H}_{ex,i} = \frac{2A_i}{\mu_0 M_{e,i}^2} \nabla^2 \mathbf{M}_i - \frac{4A_{h,i}}{\mu_0 M_{e,i} M_{e,j}} \hat{\mathbf{m}}_i \times \left( \hat{\mathbf{m}}_i \times \mathbf{M}_j \right) + \frac{A_{nh,i}}{\mu_0 M_{e,i} M_{e,j}} \nabla^2 \mathbf{M}_j \quad (i = A, B, i \neq j) \tag{22}$$

The effective field includes a number of contributions, as for the ferromagnetic model, namely demagnetizing field computed for $(\mathbf{M}_A+\mathbf{M}_B)/2$, external field, magneto-crystalline anisotropy, as well as DMI or interfacial DMI terms. The formulas given in Appendix A are now applicable to the two-sublattices separately. Temperature dependencies and further details are given in Appendix B.

**Figure 9** – (a) Antiferromagnetic resonance computed as a function of homogeneous antiferromagnetic exchange and uniaxial anisotropy, compared to the Kittel formula. (b) Antiferromagnetic spin wave dispersion computed for the same geometry in Ref. [37], with A = 5 pJ/m, $A_h$ = -20 MJ/m$^3$, $A_{nh}$ = -10 pJ/m, and $K_1$ = 50 kJ/m$^3$, using a damping factor of 0.002, compared to Equation (23) for n = 0 mode (dotted line).

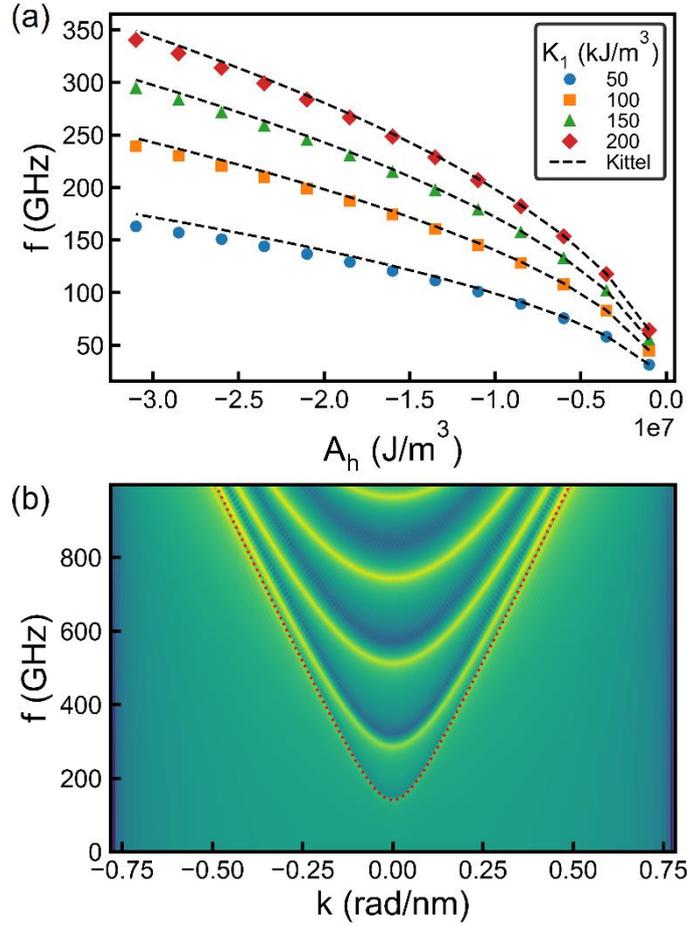

Here we test the two-sublattice model by computing the antiferromagnetic resonance (AFMR) as a function of antiferromagnetic exchange and uniaxial anisotropy. The predicted resonance frequency is given by the Kittel formula [85] as $f_0 = \sqrt{H_A(2H_E + H_A)}$ at zero external field, where $H_A = 2K_1/\mu_0 M_S$, and $H_E = 4|A_h|/\mu_0 M_s$. We compute the resonance frequency for a generic antiferromagnetic material with $M_S$ = 800 kA/m and $A$ = 13 pJ/m on each sub-lattice, as a function of homogeneous antiferromagnetic exchange and uniaxial anisotropy constant, by applying a uniform sinc pulse and taking the Fourier transform to obtain a frequency-swept AFMR peak. The excitation is applied perpendicular to the easy axis with amplitude 1 kA/m. Results are plotted in Fig 9(a), showing a good agreement with the Kittel formula over a wide range of values. We further compute the spin wave dispersion with the same method used to compute the ferromagnetic spin wave dispersion in Figure 3. Here we set A = 5 pJ/m, Ah = -20 MJ/m$^3$, K$_1$ = 50 kJ/m$^3$, and also set a nonhomogeneous exchange constant A$_{nh}$ = -10 pJ/m, with a damping constant of 0.002. The n = 0 spin wave mode analytical formula is given as:

$$w_0(k) = \gamma \sqrt{\left(2H_E + H_A + \left(\frac{2A + |A_{nh}|}{\mu_0 M_S}\right)k^2\right)\left(H_A + \left(\frac{2A + |A_{nh}|}{\mu_0 M_S}\right)k^2\right)} \tag{23}$$

The results are plotted in Figure 9(b), showing an excellent agreement with Equation (23).

The two-sublattice model in Equation (21) also includes stochastic terms, which similar to the stochastic LLB equation have zero spatial, vector components, and inter-lattice correlations, and whose components follow Gaussian distributions with zero mean and standard deviations given by:

$$H_{th,i}^{std.} = \frac{1}{\alpha_{\perp,i}} \sqrt{\frac{2k_B T(\alpha_{\perp,i} - \alpha_{\parallel,i})}{\gamma_i \mu_0 M_{S,i}^0 V \Delta t}}$$

$$\eta_{th,i}^{std.} = \sqrt{\frac{2k_B T \alpha_{\parallel,i} \gamma_i M_{S,i}^0}{\mu_0 V \Delta t}} \tag{24}$$

Similarly to the approach in Ref. [8], it can be shown the magnetization length distribution follows a Boltzmann probability distribution. For the 2-sublattice case in general this distribution is a function of the magnetization length of both sub-lattices, $m_A$ and $m_B$, and is shown below for the isotropic case (see Appendix B for further definitions).

$$P_i(m_A, m_B) \propto m_i^2 \exp\left\{-\frac{M_{S,i}^0 V}{4\mu_i m_{e,i} k_B T}\left[\frac{(m_i^2 - m_{e,i}^2)^2}{m_{e,i}}\frac{(\mu_i + 3\tau_{ij}k_B T_N \tilde{\chi}_{\parallel,j})}{2\tilde{\chi}_{\parallel i}} + \frac{(m_j^2 - m_{e,j}^2)}{m_{e,j}}3\tau_{ij}k_B T_N m_i^2\right]\right\} \tag{25}$$

$(i, j = A, B, i \neq j)$

We test this by computing a two-sublattice histogram for the magnetization length as a function of temperature, taking the generic antiferromagnetic material of Figure 9(a) with a Néel temperature $T_N = 500$ K. A temperature is set and a cubic block of antiferromagnetic material (400 nm side) with periodic boundary conditions in all directions is allowed to relax for a set time (20 ps or longer). The computed two-sublattice probability distribution is shown in Figure 10 for $T/T_N = 0.99$ as a color map. A very good agreement is obtained with the two-sublattice Boltzmann distribution from Equation (25), plotted as a wire-frame. Similar tests were repeated over a wide range of temperatures. This shows the implemented stochastic two-sublattice LLB model correctly reproduces the expected stochastic properties.

**Figure 10** – Two-sublattice antiferromagnetic magnetization length probability distribution at $T/T_N = 0.99$, showing the computed distribution as a color map, with the wire frame showing the predicted two-sublattice Boltzmann probability distribution.

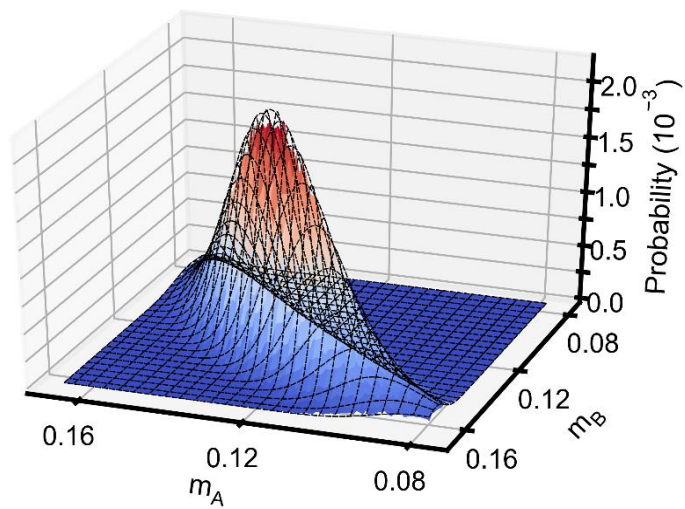

Exchange bias may also be modelled by simulating a bilayer consisting of an antiferromagnetic and a ferromagnetic mesh. The effective exchange bias field, first observed by Meikeljohn and Bean [86], coincides with the bilinear surface exchange field of Equation (5), thus exchange bias may be included by enabling the surface exchange fields at the interface between antiferromagnetic and ferromagnetic meshes [87]. By including such coupling terms between the ferromagnetic spins and one or both antiferromagnetic sub-lattices, uncompensated spins as well as compensated spins [88] may be simulated. This subject however is beyond the scope of the current work and will be addressed in a separate publication.

# 7. Benchmarking

Large-scale micromagnetics simulations require significant computational resources. An important advancement is the use of GPUs, which result in significant speed-up factors compared to CPUs [2], typically over an order of magnitude. All the computational routines in Boris may be executed on the CPU as well as on the GPU, either with single or double floating point precision. By far the most expensive term to evaluate is the demagnetizing field, which involves evaluating a convolution sum over the entire mesh – see Equation (4) – and normally takes 75% or more of the computation time in each iteration. The convolution sum may be evaluated very efficiently using the convolution theorem: a (2)3D fast Fourier transform (FFT) algorithm is used on the input magnetization; this is then multiplied with the demagnetizing kernel in the transform space, and an inverse (2)3D FFT algorithm is used to obtain the output demagnetizing field. The computational complexity of this approach increases as Nlog(N), compared to $N^2$ for the naïve evaluation of the convolution sum, thus several orders of magnitude improvement may be achieved for large number of computational cells N. The simplest method of implementing the 3D convolution algorithm consists of computing 1D FFTs along the x, y, z directions, performing the point-by-point multiplication, then computing the z, y, x inverse 1D FFTs in this order. With CUDA [30] implementations of GPU computations, these seven steps are most easily implemented using separate CUDA kernel launches. In particular the z FFTs, point-by-point multiplications, and z inverse FFTs are done in three separate steps. This can be inefficient for a small number of cells along the z direction. In Boris a new approach is taken, termed pipelined convolution, where the z (inverse)FFTs and point-by-point multiplications are done using a single CUDA kernel launch, simultaneously for all 3 vector components. This involves manually coding the FFT algorithm and results in significant performance improvement over the non-pipelined approach due to more efficient use of GPU instruction bandwidth, up to a certain number of computational cells along the z direction.

**Figure 11** – Comparison of computational performance with Mumax3, for single floating point precision CUDA computations, benchmarked on a GTX 980 Ti GPU under Ubuntu 20.04. (a) Time per evaluation as a function of total number of simulation cells for $N_z = 1$ (2D mode) and $N_z = 8$ (3D mode). (b) Speedup factor, defined as the ratio of time per evaluation as $t_{Mumax3} / t_{Boris}$, as a function of $N_z$ and total number of computational cells. In Boris, 3D computations up to 16 cells along the z direction (80 nm thickness for a 5 nm cellsize) are handled using an efficient pipelined convolution algorithm, resulting in significant speedup factors compared to non-pipelined convolution in Mumax3, up to nearly twice faster on this platform for large simulations containing over 8 million computational cells.

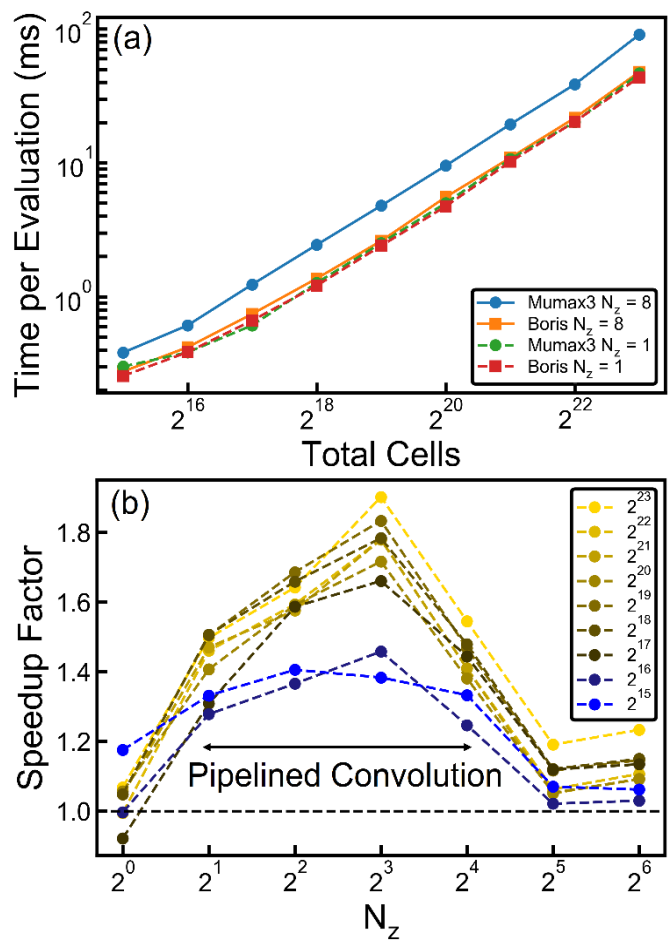

To test the efficient implementation of computational routines in Boris, benchmark comparisons with Mumax3 [2] have been performed. A testing platform consisting of a GTX 980 Ti GPU in single floating precision mode on Ubuntu 20.04 was used. An identical simulation was configured for both programs, consisting in computing the magnetization response to a perpendicular magnetic field, with effective field contributions of demagnetizing field, exchange interaction, and applied field. The RK4 evaluation method was used, and the time per evaluation was measured, noting the RK4 method consists of 4 evaluations per time step iteration. The benchmarking scripts for both programs are available in the Boris GitHub repository [28]. Typical results are shown in Figure 11(a), both for 2D and 3D modes, showing the time per evaluation as a function of total number of simulation cells N. In 2D mode the computational performance of Boris and Mumax3 are comparable, however in 3D mode Boris is found to run significantly faster. More information is obtained by plotting the speedup factor ($t_{Mumax3}$ / $t_{Boris}$) as a function of number of z cells for a wide range of total number of computational cells, shown in Figure 11(b). The pipelined convolution algorithm has been implemented up to $2^4$ cells along the z direction, thus for FFTs of up to 32 points, noting the circular convolution theorem requires doubling the input data size by zero padding when not using periodic boundary conditions. This approach is found to be significantly faster compared to Mumax3, with the speedup factor also increasing with the total number of computational cells. Above $2^4$ cells along the z direction the pipelined convolution algorithm becomes less efficient than the non-pipelined algorithm, thus $2^4$ is the largest value for which Boris implements the pipelined convolution approach, although speedup factors above 1 are still obtained for the non-pipelined convolution mode in all 3D cases.

# 8. Conclusions and Outlook

Here we've presented the main mathematical models implemented, and testing of a new comprehensive computational magnetism research software. This represents a significant addition to the body of modelling capabilities introduced in other comparable open-source software, including OOMMF [1], Mumax3 [2], and Fidimag [3]. Thus in addition to existing micromagnetics modelling software, a new multi-mesh modelling paradigm is introduced, allowing complex simulations with multiple independently discretized computational meshes and materials. This allows simulations of multi-material structures, including ferromagnetic, antiferromagnetic, ferrimagnetic, as well as non-magnetic and substrate materials, without the constraint of fitting the computations on a single uniformly discretized finite difference mesh, whilst still preserving the computational performance associated with finite difference methods. In addition to magnetization dynamics models, including LLG, LLB, as well as stochastic and two-sublattice models, Boris also implements a drift-diffusion spin transport solver in ferromagnetic materials, as well as a heat flow solver in multi-layered structures.

Whilst the implemented spin transport solver is only applicable to ferromagnetic materials, a future development consists in extending the drift-diffusion model implementation to a two-sublattice model, for example as introduced in Ref. [89], with appropriate boundary conditions. Magneto-elastic effects may be modelled in Boris either using a uniform stress, or by importing an externally computed strain or mechanical displacement, similar to the approach implemented in an OOMMF extension [90]. A future development will implement both a multi-layered elastostatics solver, as well as an elastodynamics solver [91], allowing complex simulations with non-uniform and time-dependent strains, including magneto-elastic and magnetostriction-related dynamical effects. Finally, a basic atomistic modelling [92] capability has already been introduced in Boris, with a view to implementing true multi-scale simulations [93,94] in the multi-mesh paradigm, although this was not discussed in the current work and will be treated in a separate publication.

# Appendix A – Micromagnetics Effective Field Terms

Effective field terms for various interactions are included as additive terms in $\mathbf{H}_{\text{eff}}$, either in the LLG or LLB equations. These are usually obtained from their corresponding energy density terms using the relation $\mathbf{H} = \frac{-1}{\mu_0 M_S} \frac{\partial \varepsilon}{\partial \mathbf{m}}$. The main terms implemented in Boris, not already given in the main text are shown in Table 1. Parameter definitions are not repeated here if given in the main text.

**Table 1** –Effective field terms implemented in Boris, not already given in the main text.

| Term | Formulas |
|---|---|
| Uniaxial Anisotropy | $\varepsilon = K_1 \left[1 - (\mathbf{m}.\mathbf{e}_A)^2\right] + K_2 \left[1 - (\mathbf{m}.\mathbf{e}_A)^2\right]^2$ <br><br> $\mathbf{H} = \frac{2K_1}{\mu_0 M_S}(\mathbf{m}.\mathbf{e}_A)\mathbf{e}_A + \frac{4K_2}{\mu_0 M_S}[1 - (\mathbf{m}.\mathbf{e}_A)^2](\mathbf{m}.\mathbf{e}_A)\mathbf{e}_A$ <br><br> $\mathbf{e}_A$ is the symmetry axis. |
| Cubic Anisotropy | $\varepsilon = K_1[\alpha^2\beta^2 + \alpha^2\gamma^2 + \beta^2\gamma^2] + K_2\alpha^2\beta^2\gamma^2$ <br><br> $\mathbf{H} = -\frac{2K_1}{\mu_0 M_S}[\mathbf{e}_1\alpha(\beta^2 + \gamma^2) + \mathbf{e}_2\beta(\alpha^2 + \gamma^2) + \mathbf{e}_3\gamma(\alpha^2 + \beta^2)]$ <br><br> $- \frac{2K_2}{\mu_0 M_S}[\mathbf{e}_1\alpha\beta^2\gamma^2 + \mathbf{e}_2\alpha^2\beta\gamma^2 + \mathbf{e}_3\alpha^2\beta^2\gamma]$ <br><br> $\alpha = \mathbf{m}.\mathbf{e}_1$, $\beta = \mathbf{m}.\mathbf{e}_2$, and $\gamma = \mathbf{m}.\mathbf{e}_3$, where $\mathbf{e}_3 = \mathbf{e}_1 \times \mathbf{e}_2$. |
| Direct Exchange | $\varepsilon = A\left[\left(\frac{\partial \mathbf{m}}{\partial x}\right)^2 + \left(\frac{\partial \mathbf{m}}{\partial y}\right)^2 + \left(\frac{\partial \mathbf{m}}{\partial z}\right)^2\right]$ <br><br> $\mathbf{H} = \frac{2A}{\mu_0 M_S}\nabla^2 \mathbf{m}$ <br><br> Homogeneous Neumann boundary condition: $\frac{\partial \mathbf{m}}{\partial \mathbf{n}} = 0$ |
| DM Exchange | $\varepsilon = D\mathbf{m}.(\nabla \times \mathbf{m})$ <br><br> $\mathbf{H} = -\frac{2D}{\mu_0 M_S}\nabla \times \mathbf{m}$ <br><br> The non-homogeneous Neumann boundary condition is used: <br><br> $\frac{\partial \mathbf{m}}{\partial \mathbf{n}} = \frac{D}{2A}\mathbf{n} \times \mathbf{m}$ |
| Interfacial DM Exchange | $\varepsilon = -D\mathbf{m}.((\nabla.\mathbf{m})\hat{\mathbf{z}} - \nabla m_z)$ <br><br> $\mathbf{H} = \dfrac{2D}{\mu_0 M_S}((\nabla.\mathbf{m})\hat{\mathbf{z}} - \nabla m_z)$ <br><br> For single lattice models the non-homogeneous Neumann boundary condition is used: <br><br> $\dfrac{\partial \mathbf{m}}{\partial \mathbf{n}} = \dfrac{D}{2A}(\hat{\mathbf{z}} \times \mathbf{n}) \times \mathbf{m}$ <br><br> For two-sublattice models the boundary conditions become [95]: <br><br> $\dfrac{\partial \mathbf{m}_i}{\partial \mathbf{n}} = \dfrac{D_i}{2A_i(1-c_i^2)}\left[(\hat{\mathbf{z}} \times \mathbf{n}) \times (\mathbf{m}_i - c_i\mathbf{m}_j)\right], \quad (i, j = A, B, \ i \neq j)$, <br><br> where $c_i = A_{nh}/2A_i$. |
| Demagnetizing Field | $\mathbf{H}(\mathbf{r}_0) = -\sum_{\mathbf{r}} \mathbf{N}(\mathbf{r} - \mathbf{r}_0)\mathbf{M}(\mathbf{r})$ <br><br> **N** computed using formulas in Ref. [50]. |
| Oersted | $\mathbf{H}(\mathbf{r}_0) = \sum_{\mathbf{r}} \mathbf{K}(\mathbf{r} - \mathbf{r}_0)\mathbf{J}_C(\mathbf{r})$ <br><br> **K** computed using formulas in Ref. [63] |
| Magneto-Optical | $H_{MO} = \sigma^{\pm} H_{MO}^0 f_{MO}(\mathbf{r},t)\hat{\mathbf{z}}$ |
| Roughness | $\mathbf{H}(\mathbf{r}_0) = -\left[\sum_{\mathbf{r} \in V} \mathbf{N}(\mathbf{r} - \mathbf{r}_0)G(\mathbf{r},\mathbf{r}_0)\right]\mathbf{M}(\mathbf{r}_0) \quad (\mathbf{r}_0 \in V)$ <br><br> G is computed using formulas in Ref. [27] |
| Magneto-Elastic | $\varepsilon_{mel,d} = B_1\left[(\mathbf{m}.\mathbf{e}_1)^2(\mathbf{S}_d.\mathbf{e}_1) + (\mathbf{m}.\mathbf{e}_2)^2(\mathbf{S}_d.\mathbf{e}_2) + (\mathbf{m}.\mathbf{e}_3)^2(\mathbf{S}_d.\mathbf{e}_3)\right]$ <br><br> $\varepsilon_{mel,od} = 2B_2\begin{bmatrix}(\mathbf{m}.\mathbf{e}_1)(\mathbf{m}.\mathbf{e}_2)(\mathbf{S}_{od}.\mathbf{e}_3) + (\mathbf{m}.\mathbf{e}_1)(\mathbf{m}.\mathbf{e}_3)(\mathbf{S}_{od}.\mathbf{e}_2) \\ + (\mathbf{m}.\mathbf{e}_2)(\mathbf{m}.\mathbf{e}_3)(\mathbf{S}_d.\mathbf{e}_1)\end{bmatrix}$ <br><br> $B_1$, $B_2$ are magneto-elastic constants, $\mathbf{e}_1$, $\mathbf{e}_2$, $\mathbf{e}_3$ are cubic anisotropy axes, $\mathbf{S}_d$ and $\mathbf{S}_{od}$ are diagonal and off-diagonal strain tensor terms. |

# Appendix B – Landau Lifshitz Bloch Temperature Dependences

For the two-sublattice model we introduce convenient micromagnetic parameters $\tau_i$ and $\tau_{ij} \in [0, 1]$. These are coupling parameters between exchange integrals and the phase transition temperature, such that $\tau_A + \tau_B = 1$ and $|J| = 3\tau k_B T_N$. Here $J$ is the exchange integral for intra-lattice ($i = A,B$) and inter-lattice ($i,j = A,B$, $i \neq j$) coupling respectively. For a simple antiferromagnet we have $\tau_A = \tau_B = \tau_{AB} = \tau_{BA} = 0.5$. For $\tau_A = 1$, $\tau_B = 0$ and $\tau_{AB} = \tau_{BA} = 0$, the temperature dependences given below, as well as the two-sublattice LLB equation, reduce to the ferromagnetic LLB case. Thus here we give the general case in terms of $\tau$ parameters.

The damping parameters are continuous at $T_N$ – the phase transition temperature – and given by:

$$\alpha_{\perp,i} = \alpha_i \left( 1 - \frac{T}{3(\tau_i + \tau_{ij} m_{e,j}/m_{e,i})\widetilde{T}_N} \right), \quad T < T_N$$

$$\alpha_{\parallel,i} = \alpha_i \left( \frac{2T}{3(\tau_i + \tau_{ij} m_{e,j}/m_{e,i})\widetilde{T}_N} \right), \quad T < T_N \tag{26}$$

$$\alpha_{\perp,i} = \alpha_{\parallel,i} = \frac{2T}{3T_N}, \quad T \geq T_N$$

We denote $\widetilde{T}_N$ the re-normalized transition temperature, given by:

$$\widetilde{T}_N = \frac{2T_N}{\tau_A + \tau_B + \sqrt{(\tau_A - \tau_B)^2 + 4\tau_{AB}\tau_{BA}}} \tag{27}$$

The normalised equilibrium magnetization functions $m_{e,i}$ are obtained from the Curie-Weiss law as:

$$m_{e,i} = B\left[ (m_{e,i}\tau_i + m_{e,j}\tau_{ij})3\widetilde{T}_N/T + \mu_i \mu_0 H_{ext}/k_B T \right], \tag{28}$$

where $B(x) = \coth(x) - 1/x$, and $\mu_i$ is the atomic magnetic moment.

The longitudinal relaxation field which includes both intra-lattice and inter-lattice contributions is given by:

$$\mathbf{H}_{\parallel,j} = \left\{ \frac{1}{2\mu_0 \tilde{\chi}_{\parallel,i}} \left(1 - \frac{m_i^2}{m_{e,i}^2}\right) + \frac{3\tau_{ij} k_B T_N}{2\mu_0 \mu_i} \left[ \frac{\tilde{\chi}_{\parallel,j}}{\tilde{\chi}_{\parallel,i}} \left(1 - \frac{m_i^2}{m_{e,i}^2}\right) - \frac{m_{e,j}}{m_{e,i}} (\hat{\mathbf{m}}_i \cdot \hat{\mathbf{m}}_j) \left(1 - \frac{m_j^2}{m_{e,j}^2}\right) \right] \right\} \mathbf{m}_i, \quad T < T_N$$

$$\mathbf{H}_{\parallel,j} = -\left\{ \frac{1}{\mu_0 \tilde{\chi}_{\parallel,i}} + \frac{3\tau_{ij} k_B T_N}{\mu_0 \mu_i} \left[ \frac{\tilde{\chi}_{\parallel,j}}{\tilde{\chi}_{\parallel,i}} - \frac{m_{e,j}}{m_{e,i}} (\hat{\mathbf{m}}_i \cdot \hat{\mathbf{m}}_j) \right] \right\} \mathbf{m}_i, \quad T > T_N$$

(29)

Here $\hat{\mathbf{m}}_i = \mathbf{m}_i / m_i$, and the relative longitudinal susceptibility is $\tilde{\chi}_{\parallel,i} = \chi_{\parallel,i} / \mu_0 M_{S,i}^0$, where:

$$k_B T \tilde{\chi}_{\parallel,i} = \frac{\mu_i B_i' \left(1 - 3\tau_j \tilde{T}_N B_j' / T\right) + \mu_j 3\tau_{ij} \tilde{T}_N B_i' B_j' / T}{\left(1 - 3\tau_i \tilde{T}_N B_i' / T\right)\left(1 - 3\tau_j \tilde{T}_N B_j' / T\right) - \tau_{ij} \tau_{ji} B_i' B_j' \left(3\tilde{T}_N / T\right)^2}, \quad (30)$$

and $B_i' \equiv B_{m_{e,i}}' \left[ \left(m_{e,i} \tau_i + m_{e,j} \tau_{ij}\right) 3\tilde{T}_N / T \right]$.

The equilibrium magnetization follows the temperature dependence $M_{e,i} = m_{e,i} M_{S,i}^0$. The anisotropy constant follows the temperature dependence $K_{1,i} = K_{1,i}^0 m_{e,i}^3$. The intra-lattice exchange stiffness $A_i$ has the temperature dependence $A_i = A_i^0 m_{e,i}^2$, whilst the inter-lattice exchange stiffnesses have the temperature dependences $A_{h(nh),i} = A_{h(nh),i}^0 m_{e,i} m_{e,j}$. The DMI exchange parameter follows the temperature dependence $D_i = D_i^0 m_{e,i}^2$. Note these temperature dependences can be adjusted depending on the material simulated where appropriate, for example using $m_{e,i}$ exponents computed using an atomistic model.